\documentclass[a4paper]{jpconf}
\usepackage{graphicx}
\begin{document}
\title{Quantum field theory and classical optics: determining the fine structure constant}

\author{Gerd Leuchs$^{1,2,3}$, Margaret Hawton$^{4}$ and Luis L S\'anchez-Soto$^{2,5}$}

\address{$^{1}$ Department Physik, Universit\"at Erlangen-N\"urnberg, 
Staudtstr. 7 B2,  91058 Erlangen, Germany}
\address{$^{2}$ Max Planck Institute for the Science of Light,
Staudtstr. 2, 91058 Erlangen, Germany}
\address{$^{3}$ Department of Physics, University of Ottawa, 
25 Tempelton Street,  Ottawa, ON, K1N 6N5, Canada}
\address{$^{4}$ Department of Physics, Lakehead University, 
955 Oliver Road, Thunder Bay, ON, P7B 5E1}
\address{$^{5}$ Departamento de \'Optica, Facultad de F\'{\i}sica, 
Universidad Complutense, 28040~Madrid,  Spain}
\ead{gerd.leuchs@mpl.mpg.de}

\begin{abstract}
  The properties of the vacuum are described by quantum physics
  including the response to external fields such as electromagnetic
  radiation. Of the two parameters that govern the details of the
  electromagnetic field dynamics in vacuum, one is fixed by the
  requirement of Lorentz invariance
  $c= 1/\sqrt{\varepsilon_{0} \mu_{0}}$. The other one,
  $Z_{0}= \sqrt{\mu_{0}/\varepsilon_{0}} = 1/(c\varepsilon_{0})$ and
  its relation to the quantum vacuum, is discussed in this
  contribution. Deriving $\varepsilon_{0}$ from the properties of the
  quantum vacuum implies the derivation of the fine structure
  constant.
\end{abstract}

\section{Introduction}

At the starting point of the prototype quantum field theory, quantum
electrodynamics (QED), are the equations of Maxwell and
Dirac. Formulation of a consistent theory from scratch was delayed by
the enormous difficulties encountered, until Feynman and Schwinger
decided to postulate Maxwell's equations in vacuum and build the
theory from there by adding the interaction between the
electromagnetic field and the electron. When Maxwell formulated the
equations carrying his name, the mathematical structure was motivated
by the attempt to bring all laws of electromagnetism known hitherto
into one consistent form. There are, however, parameters in Maxwell's
equations, which had to be determined by experiment. These two
parameters are, depending on your choice of the unit system:
$\varepsilon_{0}$ and $\mu_{0}$ or the speed of light in vacuum and
the impedance of the vacuum, or whatever the choice of units is. The
compatibility with special relativity requires Maxwell's equation to
be Lorentz-invariant, which means that the speed of light in vacuum
and the limiting speed in special relativity are identical. This
related one parameter to the rest of physics,
$c= 1/\sqrt{\varepsilon_{0} \mu_{0}}$, 
while the other parameter, the vacuum impedance
$Z_{0}= \sqrt{\mu_{0}/\varepsilon_{0}}= 1/(c\varepsilon_{0})$ was
still an experimental parameter unrelated to any other area in
physics. This we challenge by postulating that this second parameter
is intimately related to the property of the modern quantum vacuum. As
will become clear, this also determines the fine structure constant.

\section{A dielectric model for the quantum vacuum}

In ultra high electric fields the vacuum is predicted to break down
forming an electron--positron plasma. At lower field strength, virtual
electron--positron pairs are polarized forming virtual ephemeral
electric dipoles. This vacuum polarization provides the partial
shielding of point charges.  In analogy to
the treatment of a dielectric medium, this vacuum polarization caused
by a light field should appear in Maxwell's equations. The quantity to
look for is the electric displacement field introduced by
Maxwell. This is the first term on the right hand side of
\begin{equation}
  \label{eq:1}
 \mathbf{D} =  \varepsilon_{0} \mathbf{E}  +   \mathbf{P} \, .
\end{equation}
One might be tempted, as we are, to think of
$ \varepsilon_{0} \mathbf{E}$ as of the polarization of the
vacuum. The vacuum would then be a modern version of the aether, with
all the quantum properties and with relativistic symmetry~\cite{Laughlin:2005aa}.

Along this line the response of the vacuum to an external electric
field is the polarization of virtual particle--antiparticle pairs
ubiquitous in the vacuum. And the precision with which Maxwell's
equations describe low energy electro-magnetic experiments suggests
that they already contain the linear response of the vacuum. This
linear response can be calculated on the back of an envelope using a
simple model~\cite{Leuchs:2010aa,Leuchs:2013aa}. There it is argued
that the dipole moment induced in a single virtual pair is inversely
proportional to the third power of the rest mass of the particles. The
polarization in equation~(\ref{eq:1}) is a dipole moment density so we
need to divide the induced dipole moment by the volume, a single
particle pair occupies. A first guess for this volume is this
particle's Compton wavelength cubed, which corresponds to a momentum
cut-off of $p=mc$ when integrating over all momenta. In the resulting
expression for the polarization of the vacuum due to virtual
electron-positron pairs the mass of the electron drops out. The
surprising consequence is that the electronic contribution depends
only on the square of the electron charge and not on the mass. The
same must then be true for all other types of elementary
particles. They all contribute to the vacuum polarization in a similar
way proportional to the square of their charge and one thus has to sum
over all different types of elementary particles. Therefore, the
parameters used in Maxwell's equations, determined experimentally so
far, are related to the relativistic quantum vacuum and allow one to
deduce information about all charged elementary particles known and
unknown. According to this simple model the fine structure constant is
given by the sum over all types of elementary particles. For
$\varepsilon_{0}$ one obtains
\begin{equation}
  \label{eq:2}
  \varepsilon_{0} = \frac{1}{\hbar c} \, f
  \sum_{j}^{\mathrm{e. \, p.}} q_{j}^{2} \, . 
\end{equation}
The factor $f$ in our model is of order unity, but depends on the
details of renormalization. The index runs over all different
types of charged elementary particle--antiparticle pairs (e. p.):
$j=1$ electron, $j=2$ muon, $j=3$ tauon, $j=4$ to 21 the different
versions of quarks, $j=22$ the W$^{+}$ boson and what else is out
there ($j>22 ?$).

\section{Implications of the result of model}

The charge of the electron, which we see in low energy experiments, is
to our current understanding the partially screened bare point
charge. The screening is due to the polarization of the vacuum. The
contribution to the point charge screening by a particular type of
elementary particle happens at a distance from the point charge
roughly equivalent to the Compton wavelength of the particular type of
particles. The same holds for the other types of elementary
particles~\cite{Gottfried:1986aa}. As is well known from high-energy
scattering experiments, for example between two electrons, the
apparent charge of the electron increases as the scattering particles
come closer. As a consequence, the screening by vacuum polarization is
reduced when $(\omega_{\mathbf{k}}/c)^{2}$ of the photon exchanged by
the scattering particles differs more and more from
$\mathbf{k}^{2}$. This is quantified by the off-shellness
$k^{2}=\omega _{\mathbf{k}}^{2}/c^{2}-\mathbf{k}^{2}$.  Equivalently,
the fine structure constant, which is proportional to $e^{2}$ likewise
increases. In our model, however, it is the dielectric permittivity,
which decreases as you start to leave out the low rest mass or high
Compton wavelength particles contributing to the screening at larger
distances. Inserting equation (\ref{eq:2}) into
\begin{equation}
  \label{eq:3}
  \alpha = \frac{1}{4 \pi \varepsilon_{0}} \frac{e^{2}}{\hbar c} \,,
\end{equation}
yields
\begin{equation}
  \label{eq:4}
  \alpha^{-1} = 4 \pi f  \sum_{j}^{\mathrm{e. \, p.}} 
 \left (  \frac{q_{j}}{e} \right )^{2} \, .
\end{equation}
This can be compared with the running fine structure constant routinely used
in particle physics~\cite{Peskin:1995aa,Hogan:2000aa}
\begin{equation}
  \label{eq:5}
\alpha ^{-1} ( k^{2} ) \simeq  \alpha ^{-1}(0)-
\frac{1}{3\pi}  \sum_{j}^{\mathrm{e. \, p.}}   
 \left (  \frac{q_{j}}{e} \right )^{2}
\ln \left( \frac{\hbar ^{2}k^{2}}{m_{j}^{2}c^{2}}\right) \, ,
\end{equation}
where  $k^{2}=\omega _{\mathbf{k}}^{2}/c^{2}-\mathbf{k}^{2}$ and the
limit $\hbar ^{2} | k^{2} | \gg m_{j}^{2}c^{2}$ is assumed.
The value of $k^{2}$ for which the right-hand side of
equation~(\ref{eq:5}) vanishes, $k^{2} = \Lambda_{L}^{2}$, is referred to as the Landau
pole~\cite{Landau:1954aa}. So, 
equation~(\ref{eq:5}) gives
\begin{equation}
\alpha ^{-1}(0)=\frac{1}{3\pi } 
\sum_{j}^{\mathrm{e.\,p.}} \left (  \frac{q_{j}}{e} \right )^{2} 
\ln \left( \frac{\hbar ^{2}\Lambda_{L}^{2}}{m_{j}^{2}c^{2}}%
\right)\,.  
\label{eq:6}
\end{equation}
It is obvious that all different types of virtual elementary particles
contribute to $\alpha$. This looks very much like equation~(\ref{eq:4})
derived from the simple model, if one replaces the fudge factor $f$ by 
\begin{equation}
  \label{eq:7}
  f = \frac{1}{12 \pi^{2}} 
\left \langle   \ln \left (
   \frac{\hbar^{2}\Lambda_{L}^{2}}{m_{j}^{2} c^{2}} \right )  \right \rangle_{j}  \, .
\end{equation}
Note that for the electron and the $W^{+}$ boson, the logarithmic term
inside the angle brackets varies
only by 23\%: from 101 to 78, if one chooses $\hbar \Lambda_{L}/c$ be
the Planck mass. The averaged result $f \simeq 0.7$ can be factored out of the
sum. The relevance of the cut-off being determined by the Planck mass
is that then
$\alpha^{-1}(k^{2} \rightarrow \infty) = \varepsilon (k^{2}
\rightarrow \infty) = 0$ in the bare vacuum.  It is worth noting that
in the beginning of quantum field theory several groups postulated a
formula very similar to equations~(\ref{eq:4}) and
(\ref{eq:5}). However, this was before quarks were discovered at a
time when one assumed that all charged elementary particles had the
same electric charge $e$. In this case the sum simply turns into the
number $\nu$ of different types of elementary
particles~\cite{Landau:1955aa,Zeldovich:1967aa}
\begin{equation}
  \label{eq:8}
  \alpha^{-1} = \frac{1}{3 \pi} \nu \ln 
  \left ( \frac{\hbar c}{G m^{2}} \right ) \, .
\end{equation}
The main result here is the closed form in equation~(\ref{eq:6}) for the
fine structure constant, derived in a fully quantum field theoretical
treatment and inspired by the simple dielectric model of the vacuum
above.

\bigskip

\providecommand{\newblock}{}

\end{document}